\documentclass{frontiersSCNS}

\usepackage[english]{babel}
\usepackage{amsmath}
\usepackage{graphicx}

\usepackage{url,hyperref,lineno,microtype}
\usepackage[onehalfspacing]{setspace}
\def\keyFont{\fontsize{8}{11}\helveticabold }
\def\firstAuthorLast{Aledavood {et~al.}} %use et al only if is more than 1 author
\def\Authors{T. Aledavood\,$^{1,*}$, S. Lehmann\,$^{2,3}$ and J. Saram\"aki\,$^1$}
\def\Address{$^{1}$Department of Computer Science, P.O. Box 15400, Aalto University School of Science, FI-00076 AALTO, Finland \\
$^{2}$DTU Compute, Technical University of Denmark, Kgs. Lyngby, Denmark\\
$^{3}$The Niels Bohr Institute, University of Copenhagen, Copenhagen, Denmark}

\def\corrAuthor{Talayeh Aledavood}
\def\corrAddress{Department of Computer Science, P.O. box 15400, Aalto University School of Science, FI-00076 AALTO, Finland}
\def\corrEmail{talayeh.aledavood@aalto.fi}

\begin{document}
\onecolumn
\firstpage{1}

\title[On the Digital Daily Cycles of Individuals]{On the Digital Daily Cycles of Individuals} 

\author[\firstAuthorLast ]{\Authors} %This field will be automatically populated
\address{} %This field will be automatically populated
\correspondance{} %This field will be automatically populated

\extraAuth{}% If there are more than 1 corresponding author, comment this line and uncomment the next one.
%\extraAuth{corresponding Author2 \\ Laboratory X2, Institute X2, Department X2, Organization X2, Street X2, City X2 , State XX2 (only USA, Canada and Australia), Zip Code2, X2 Country X2, email2@uni2.edu}

\maketitle

%%%%%%%%%%%%%%%%%%%%%%%%%%%%%%%%%%%%%%%%%%%%%%%%%%%%%%%%%%%%%%%%%%%%%%%%%%%%%%%%%%%%%%%%%%%%%%%%%%%%%%%%%%%%%%%%%%%%%%%%%%%%%%%%%%%%%%%%%%%%%%%%%%%%%%%%%%%%%%%%%%%%%%%%%%%%%%%%%%%%%%%%%%%%%%%%%%%%%%%%%%%%%%%%%%%%%%%%%%%%%%%%%%%%%%%
%%% The sections below are for reference only.
%%%
%%% For Original Research Articles, Clinical Trial Articles, and Technology Reports the section headings should be those appropriate for your field and the research itself. It is recommended to organize your manuscript in the
%%% following sections or their equivalents for your field:
%%% Abstract, Introduction, Material and Methods, Results, and Discussion.
%%% Please note that the Material and Methods section can be placed in any of the following ways: before Results, before Discussion or after Discussion.
%%%
%%%For information about Clinical Trial Registration, please go to http://www.frontiersin.org/about/AuthorGuidelines#ClinicalTrialRegistration
%%%
%%% For Clinical Case Studies the following sections are mandatory: Abstract, Introduction, Background, Discussion, and Concluding Remarks.
%%%
%%% For all other article types there are no mandatory sections.
%%%%%%%%%%%%%%%%%%%%%%%%%%%%%%%%%%%%%%%%%%%%%%%%%%%%%%%%%%%%%%%%%%%%%%%%%%%%%%%%%%%%%%%%%%%%%%%%%%%%%%%%%%%%%%%%%%%%%%%%%%%%%%%%%%%%%%%%%%%%%%%%%%%%%%%%%%%%%%%%%%%%%%%%%%%%%%%%%%%%%%%%%%%%%%%%%%%%%%%%%%%%%%%%%%%%%%%%%%%%%%%%%%%%%%%

\begin{abstract}
Humans, like almost all animals, are phase-locked to the diurnal cycle. Most of us sleep at night and are active through the day. Because we have evolved to function with this cycle, the circadian rhythm is deeply ingrained and even detectable at the biochemical level. However, within the broader day-night pattern, there are individual differences: e.g., some of us are intrinsically morning-active, while others prefer evenings. In this article, we look at digital daily cycles: circadian patterns of activity viewed through the lens of auto-recorded data of communication and online activity. We begin at the aggregate level, discuss earlier results, and illustrate differences between population-level daily rhythms in different media. Then we move on to the individual level, and show that there is a strong individual-level variation beyond averages: individuals typically have their distinctive daily pattern that persists in time. We conclude by discussing the driving forces behind these signature daily patterns, from personal traits (morningness/eveningness) to variation in activity level and external constraints, and outline possibilities for future research.

%%% Leave the Abstract empty if your article falls under any of the following categories: Editorial Book Review, Commentary, Field Grand Challenge, Opinion or specialty Grand Challenge.
\section{}
%As a primary goal, the abstract should render the general significance and conceptual advance of the work clearly accessible to a broad readership. References should not be cited in the abstract.

\tiny
 \keyFont{ \section{Keywords:} circadian rhythms, electronic communication records, mobile phones, digital phenotyping, individual differences} %All article types: you may provide up to 8 keywords; at least 5 are mandatory.
\end{abstract}

\section{Introduction}

Almost all life on Earth is affected by the planet's 24-hour period of rotation. Humans are no different; the rhythms of our lives are phase-locked with the diurnal cycle. Because our bodies have evolved to cope with the external environment, we have genetic circadian pacemaker circuits that intrinsically follow a period of approximately 24 hours (the circadian period length may vary from one person to another, vary by age and there are known gender differences~\citep{schmidt2012, Duffy2011}). The operation of these circadian circuits manifests at various levels: biochemical, physiological, psychological, and in various markers from hormone levels to body temperature~\citep{Kerkhof1985,Czeisler1999,Panda2002,Baehr2000}. While our daily rhythms can be modulated by exogenous factors (e.g.~decoupling alertness from the sleep/wake cycle~\citep{Folkard1985}), there is a very strong endogenous component in these rhythms, as indicated by the persistence of a near-24 hour rhythm in the absence of environmental cues or despite imposition of a non-24 hour schedule~\citep{Kleitman1963,Wever1979}. 

Within this broader pattern, however, there are substantial inter-individual differences. Such differences are apparent in the existence of \emph{chronotypes} -- morning types and evening types, those who go to bed early and those who find it difficult to wake up early. The traits of \emph{morningness} and \emph{eveningness} correlate with distinctive temporal patterns of physiological and psychological variables, such as body temperature and efficiency. They also appear to be linked to gender as well as personality traits; in particular, studies have shown weak negative correlations of morningness with extraversion and sociability~\citep{Tsaousis2010,Keren2010}.

The daily rhythms that humans follow are  visible in the digital records that are left in the wake of human online activity. Population-level and system-level daily rhythms can be observed in time variation of activity in Youtube, Twitter and Slashdot, and in frequency of edits in Wikipedia and OpenStreetMap~\citep{Gill2007,Kaltenbrunner2008,Yasseri2012,Yasseri2013}.
They are also seen in the frequency of mobile telephone calls~{\citep{Jo2012,Krings2012}}, and in traces of human mobility derived from mobile phone data~\citep{Song2010,Ahas2010,louail2014}. But what do the circadian patterns displayed by activity levels in an online system actually reveal about human behaviour? The behaviour of an online system is determined by a number of factors: the day/night cycle, the function and purpose of the system in question (e.g.~work-related emails mostly being sent during office hours, see below), the variation of behaviours of user groups (e.g.~Wikipedia edits from multiple time zones), and, importantly, variation at the individual level.

In this paper, we discuss findings regarding the daily patterns in electronic records of human communication, along with results of analyses that illustrate such patterns in four different datasets. We start at the aggregate level, studying system-level average patterns and discuss the origins of the findings. From the system level, we will move on to the level of individuals, and focus on the variation that remains hidden within system-level averages: individual differences reflected in persistent, distinct daily activity patterns. This part confirms that earlier findings of persistent individual differences in a mobile telephone dataset~\citep{Aledavood2015} are general, and that persistent, distinct daily patterns of individuals are common to different communication channels. We conclude by discussing the implications of these findings, and address future research questions from large-scale analysis of sleep habits of individuals with big data to daily activity patterns as part of digital phenotypes.

\section{Daily patterns at the aggregate level}

\subsection{Previous work}
Let us begin by discussing observations of digital daily cycles in different systems at the aggregate level, computed from digital records of communication and online activity. In every instance where the temporal variation of the activity levels in such systems is monitored, the result is a periodic pattern of activity on several time scales~\citep{minireview}. The longest scale is that of a calendar year, where special periods such as holidays can typically be distinguished (see, e.g.,~\cite{Krings2012}). Then there is a weekly cycle, where weekends typically differ from weekdays, and where there can be differences between weekdays as well~\citep{Gill2007,Kaltenbrunner2008,Krings2012,Yasseri2012,Vajna2012}. Finally, there is a daily pattern which may significantly differ between different systems.

We stress that any observed system-level pattern rises out of the superposition of a multitude of individual patterns, and attributing system-level behaviour to individuals would amount to an ecological fallacy. Therefore, interpreting what the system-level patterns represent remains a non-trivial task. Solving the problem of disentangling the superposition of daily patterns, however, may provide important information of the user population. A good example of this is ~\citep{Yasseri2012}, where the authors studied Wikipedia in various languages, and were able to infer the geographical spread of their editor base from the assumption that the observed edit frequency cycles are a superposition of circadian patterns on different time zones. The method is based on the argument that Wikipedias in different languages exhibit universal daily patterns, with minima and maxima at around the same time of the day (when correcting for time zones). 

Temporal patterns of activity have been studied for different online platforms. For example, in ~\cite{Yasseri2013}, the authors look at differences between editing patterns on OpenStreetMap, which is a geo-wiki, for two different cities (London and Rome). Circadian patterns of edits for the two cities have been compared to each other and to that of Wikipedia edits. The authors also followed changes in the circadian rhythms for each of the two cities over several years. In ~\cite{Thij2014}, daily and weekly patterns of Twitter activity in different languages have been studied and it has been shown that circadian patterns emerge for tweets in all the studied languages. In ~\cite{noulas2011}, the authors have looked at data from Foursquare and found geo-temporal rhythms in activity both for weekdays and weekends.  

Analysis of aggregate-level daily cycles with geospatial information has been used in the context of cities and transport. As an example, in~\cite {toole2012}, the authors infer dynamic land use of different parts of a city based on temporal patterns of mobile phone activity in different locations. In ~\cite{Ahas2010}, temporal data is combined with location data from mobile phones. Comparing daily rhythms for different days of the week, the authors show a significant difference in mobility of suburban commuters  in city of Tallinn on weekends as compared to work days. In ~\cite{louail2014}, the authors investigate the daily rhythms of different Spanish cities in terms of spatiotemporal patterns of mobile phone usage, and show how the structure of hotspots, places of frequent usage, allow them to distinguish between different cities. In ~\cite{dong2015}, Call Detail Record (CDR) data for a period of 5 months from Cote d'Ivoire is used to detect unusual crowd events and gatherings.  

As a more applied and non-conventional example of the analysis of daily rhythms, in May 2014 a number of different news outlets (\emph{e.g.,} ~\cite{Bloomberg2014}) described how an elaborate campaign run by Iranian hackers on social media, targeting American officials and figures, was revealed only after analysing the temporal patterns of three years of activity. The daily and weekly activity patterns of the hackers matched precisely the activity profile of Tehran (i.e.~low activity at lunch hours of Tehran local time, and little or no activity on Thursdays and Fridays which are weekend days in Iran).  

Finally, let us mention that electronic records contain evidence of daily/weekly patterns that go beyond activity rates. Using network analysis \cite{Krings2012} show that when mobile telephone calls between individuals are aggregated to form networks, the structural features of those networks differ depending on the starting time of the aggregation process. In particular, weekends differ from weekdays. It is probable that the explanation is that during weekends, communication is mainly targeted to close friends and relatives who reside within the dense core of one's egocentric network. At a smaller scale, in ~\cite{Aledavood2015}, the authors show that closest friends are frequently called in the evenings.

\subsection{Results}

In this work, we study three different datasets, one with calls, one with calls and text messages, and one containing email records~\citep{eckmann2004}. For calls, we use the Reality Mining dataset~\citep{eagle2006} and another mobile phone dataset containing data from a small town in a European country with a population of around ~8000 people, a subset of the data used in \emph{e.g.}~\cite{Karsai2011}). For the latter, we also study text messages. For all sets, we use 8-week slices. A summary of different sets can be found in Table~\ref{table_data}. Preprocessing of the data is discussed in Methods.

\begin{table}[h!]\textbf{\refstepcounter{table}\label{table_data} Table \arabic{table}.}{ Overview of the datasets used in this study. }

\centering
\processtable{}
 {\begin{tabular}{|l| c |c |c|}
 \hline
 \multicolumn{1}{|c|}{Dataset} & Participants & Active Users & Total Events \\ \hline \hline 
 & & & \\ [-1.5ex]
 Reality Mining Call & 87 & 47  & 14,187 \\
 Town Call & 1204 & 277 & 45,844\\
 Town Text & 708 & 64 &  13,014 \\
 Email & 2430 & 431 & 206,723 \\ \hline
 \end{tabular}}{}
\end{table}

\begin{figure*}[!ht]
\includegraphics[width=0.9\linewidth]{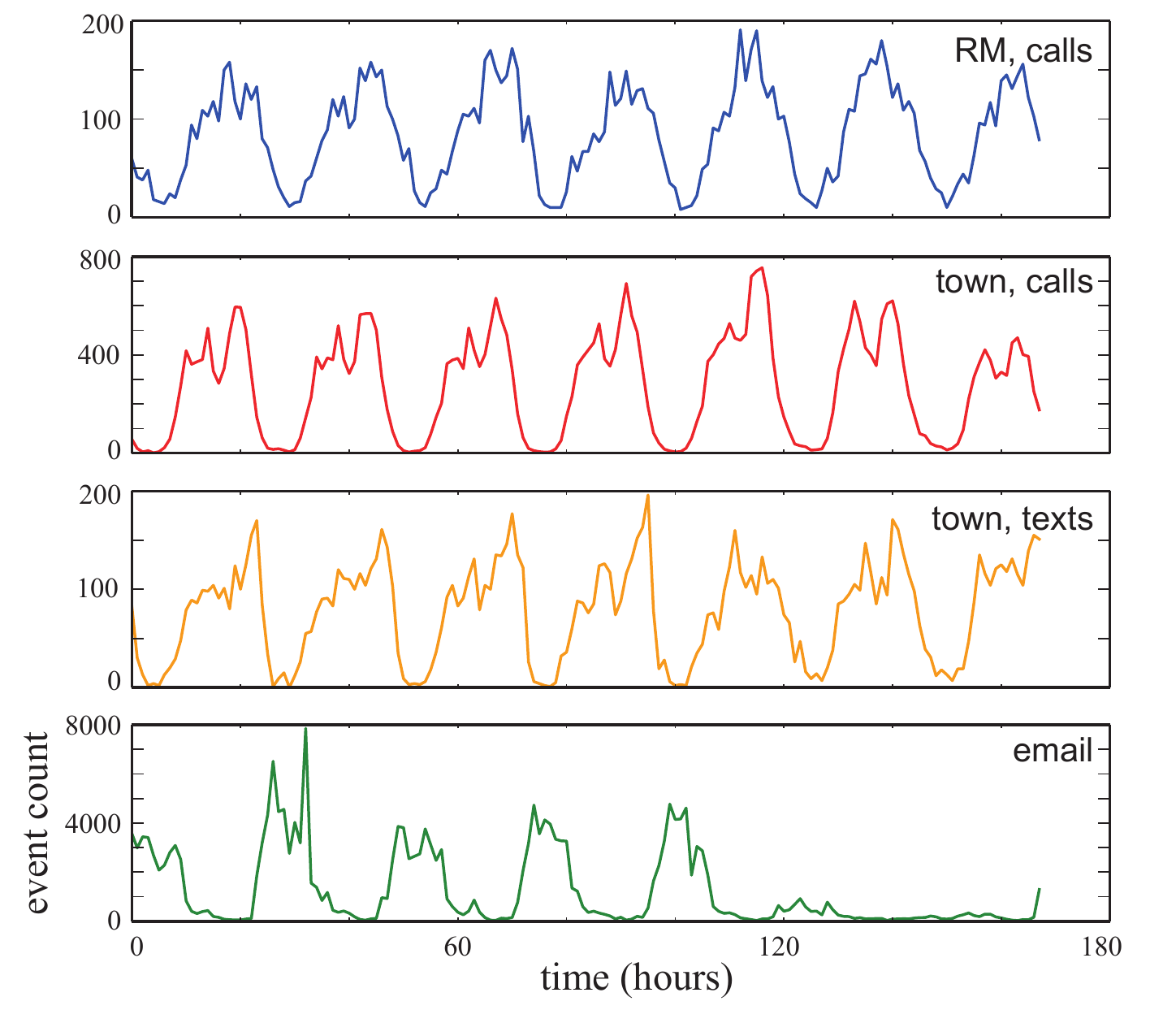}
\caption{Number of events per hour for each day of week in our datasets. This curve has been aggregated over the entire 8-week period. From top to bottom: calls in Reality Mining, calls and texts in small town, and emails. We observe strong diurnal patterns in all  datasets; for the small town datasets there are also differences between calls and and texts activity. The email dataset shows decreased activity during weekends.}
\label{fig:aggregated}
\end{figure*}

As the first step, we look at aggregated hourly event frequencies for each of the four different sets (Fig.~\ref{fig:aggregated}). It is clear that while the sleep/wake cycle is apparent in each set, there are also noticeable differences. Calls in the European town show a double-peaked daily curve, whereas the Reality Mining data displays no such pattern. It is possible that this is due to different conventions; students in Boston can be expected to behave differently than people in a small European town. Note that for the Reality Mining data, time zone information is not available, so we have manually shifted them such that the lowest points correspond to night and there is a possibility that this estimate is inaccurate. However, this only affects the phase of the pattern, not its shape.

Interestingly, in both call datasets, the highest peak occurs on the fifth day (Friday). Also note the very low email activity level during the weekend in the email data. For email, time stamps are relative to some unknown $t_0$, so the daily cycles appear shifted compared to the other datasets.

\begin{figure*}[t]
\includegraphics[width=0.75\linewidth]{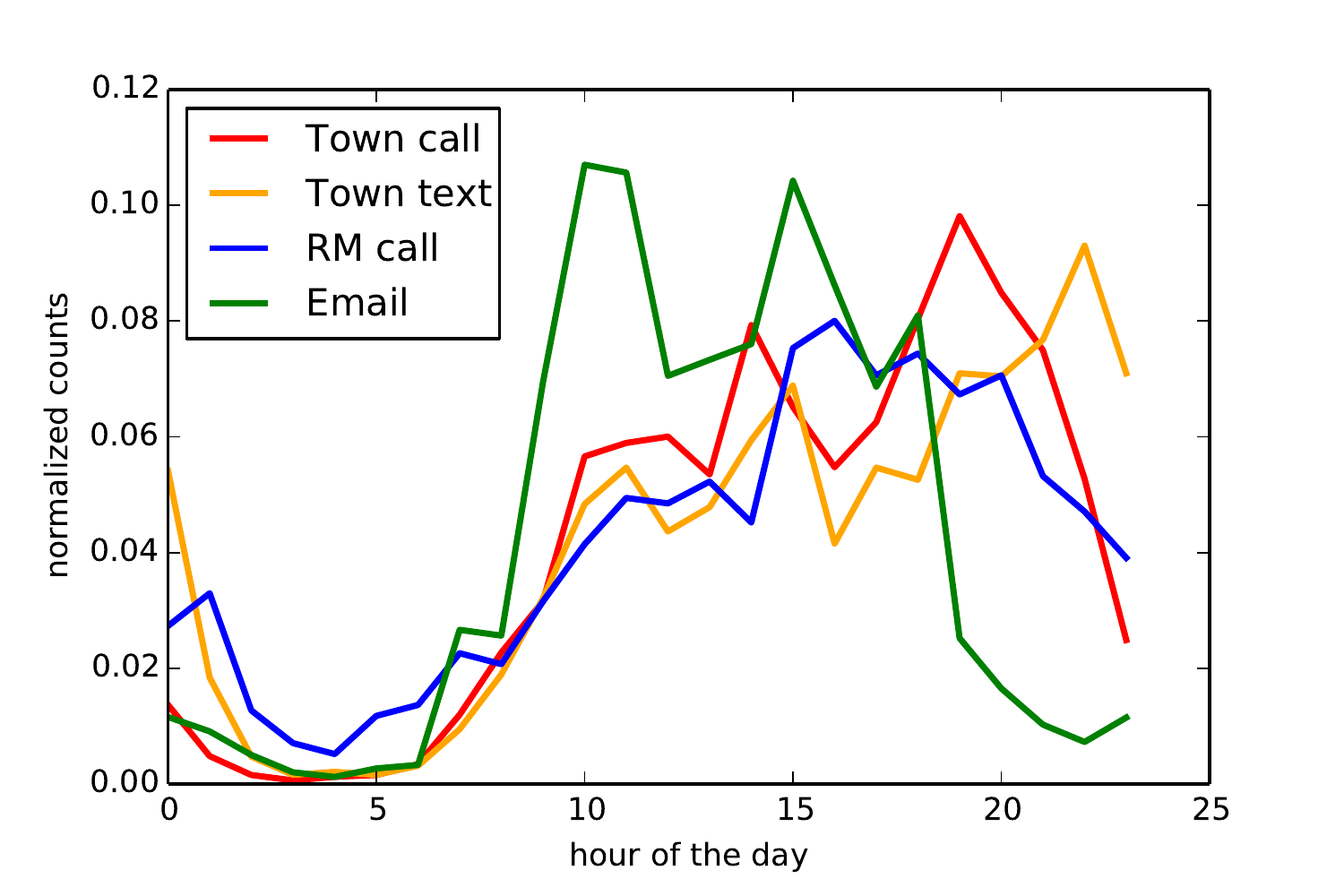}
\caption{The daily pattern in each of the datasets, computed as an average over all Wednesdays in the data. Colours are the same as in Fig.~\ref{fig:aggregated}. We observe distinct patterns across the various data channels. Email activity is early in the day, whereas (unobtrusive) text messages peak late at night.}
\label{fig:wednesday}
\end{figure*}

In Fig.~\ref{fig:wednesday} we focus on the difference between daily cycles the various datasets. Here, we plot the average daily patterns in each system on the third day of the week.
%, but could be another day if weekend is not on Saturday and Sunday in the source place of the data) . 
Since there is no exact timezone information for Reality Mining and email datasets, we identified the third day of the week by assuming that two low-activity days correspond to the weekend. We also aligned the timelines by assuming that the lowest activity of the day occurs at 4 AM for all datasets. We then average over the third-day patterns across all eight weeks in each set.
As in~\cite{Aledavood2015_2}, we find differences between the communication channels: for the small town dataset, the peak of text messages is later than that of calls. This is perhaps due to different nature of these channels; while getting calls in the late hours might not be appreciated, receiving text messages which are much less obtrusive is still acceptable.

\section{Daily patterns at the level of individuals}

\subsection{Previous work}
In~\citep{Aledavood2015}, two present authors investigated individual-level daily cycles in mobile phone call data from 24 individuals (12 male and 12 female) over 18 months. The data collection was performed in a setting where the participants completed high school some months after the collection began, and then started their first year at university, often in another city, or went to work. This design guaranteed a high turnover in their social networks~\citep{Saramaki2014}, and provided an opportunity to study a major change in their life circumstances. Looking at individual-level daily call patterns, however, it was clear that there were persistent individual differences; each individual has their distinctive daily cycle despite social network turnover and changes in circumstances. This observation speaks in favour of intrinsic factors (such as the aforementioned chronotypes) dominating individual-level variations in daily patterns (see Discussion).

%Having access to extensive survey in addition to communication data of the individuals, we knew that all individuals finished high-school a few month after begining of data collection and later moved to another city or stayed in the same place and either continued their studies, started to work or took a year off. These individuals, have a high turnover in their social network~\citep{Saramaki2014}, however looking at individul daily patterns for each ego, separately for three 6-month time intervals, we see that individuals show persistent daily patterns. To have a refrence, we calculate distance (Jensen-Shannon divergence) of daily patterns of each individual and all the other ones in the same time interval ($d_{self}$) and compare this to distance of daily patterns of each individual in different time intervals ($d_{ref}$). We show that on average $d_{self}$ is smaller than $d_{ref}$  

\subsection{Results}
Continuing the analysis of the four datasets, we first calculate for each set the  daily patterns for each individual (``ego'') by counting the total number of events associated with the ego at each hour of the day through the whole 8 weeks. The counts are then normalised to one for each ego to yield that person's daily activity pattern. As a reference, we also compute the average pattern over all egos from the normalised patterns. 

\begin{figure*}[t]
\includegraphics[width=0.9\linewidth]{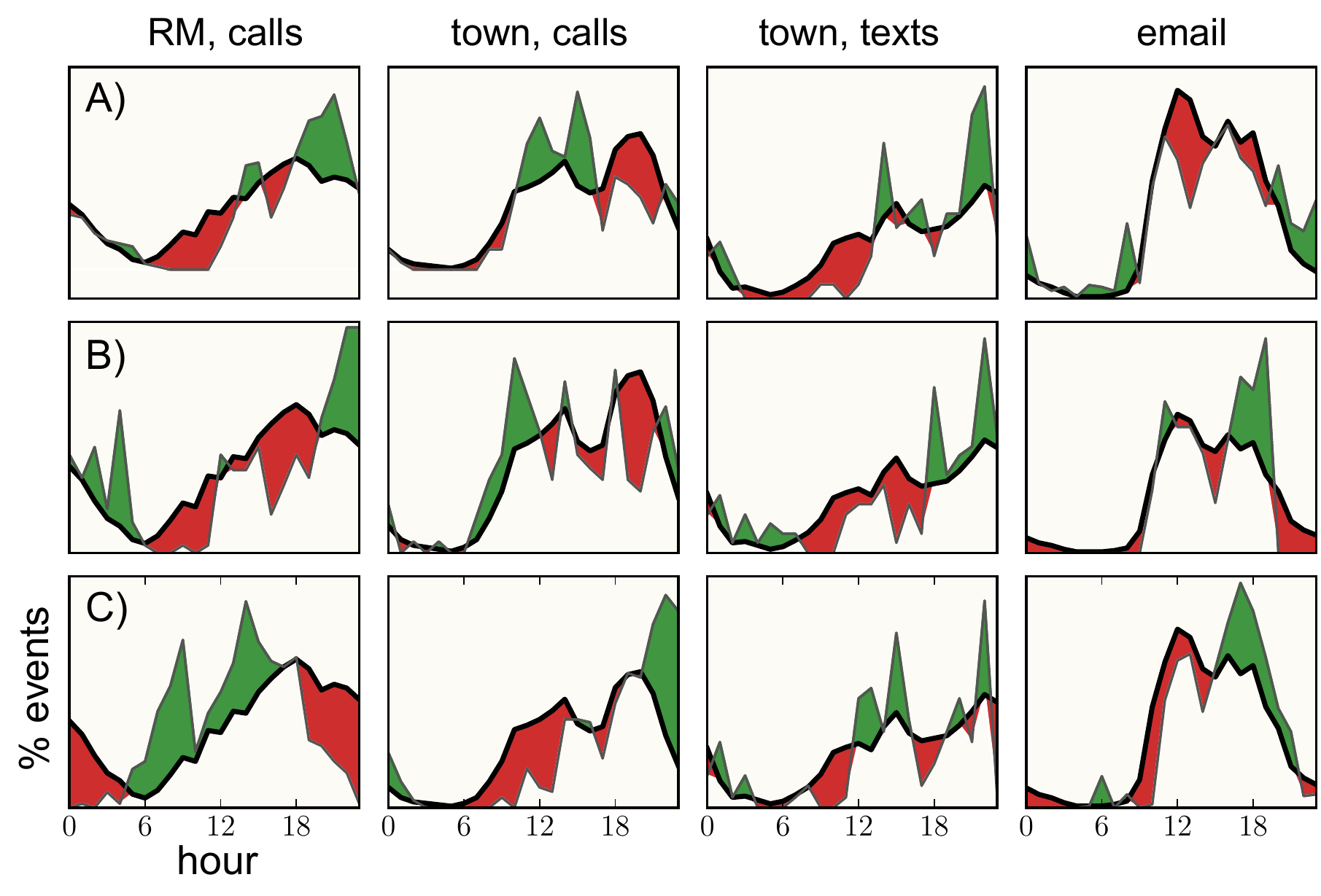}
\caption{A sample of 12 individual-level daily patterns for four datasets to illustrate the diverse nature of individual patterns. Columns correspond to datasets, while rows A)-C) correspond to different individuals (who are not the same across datasets). The black line shows the average daily pattern for the dataset in question---and therefore is the same in each column---whereas green/red areas denote where this individual's pattern is above or below average. We observe that in almost every case, the individual patterns differ strongly from the average behaviour, for example by increased calling frequency during mornings, mid-days, or evenings.}
\label{fig:average_daily_patterns}
\end{figure*}

Fig.~\ref{fig:average_daily_patterns} displays a sample of the individual-level daily patterns for each dataset. For each set, we have picked three egos to demonstrate individual differences; for each ego, their differences from the aggregated average are emphasized by red and green colours. For all datasets, we can observe clear variation between individuals.
Considering the differences between the aggregate and individual daily cycles serves two purposes. While the average pattern in each dataset reveals general underlying mechanisms, the individual patterns show that each person has their own preferences for the timing of communication with others.
The daily communication cycles point at variation beyond morningness and eveningness: while individuals clearly have different sleep/wake cycles, they also have their specific patterns during their wakefulness periods. 

\begin{figure*}[t]
\includegraphics[width=0.9\linewidth]{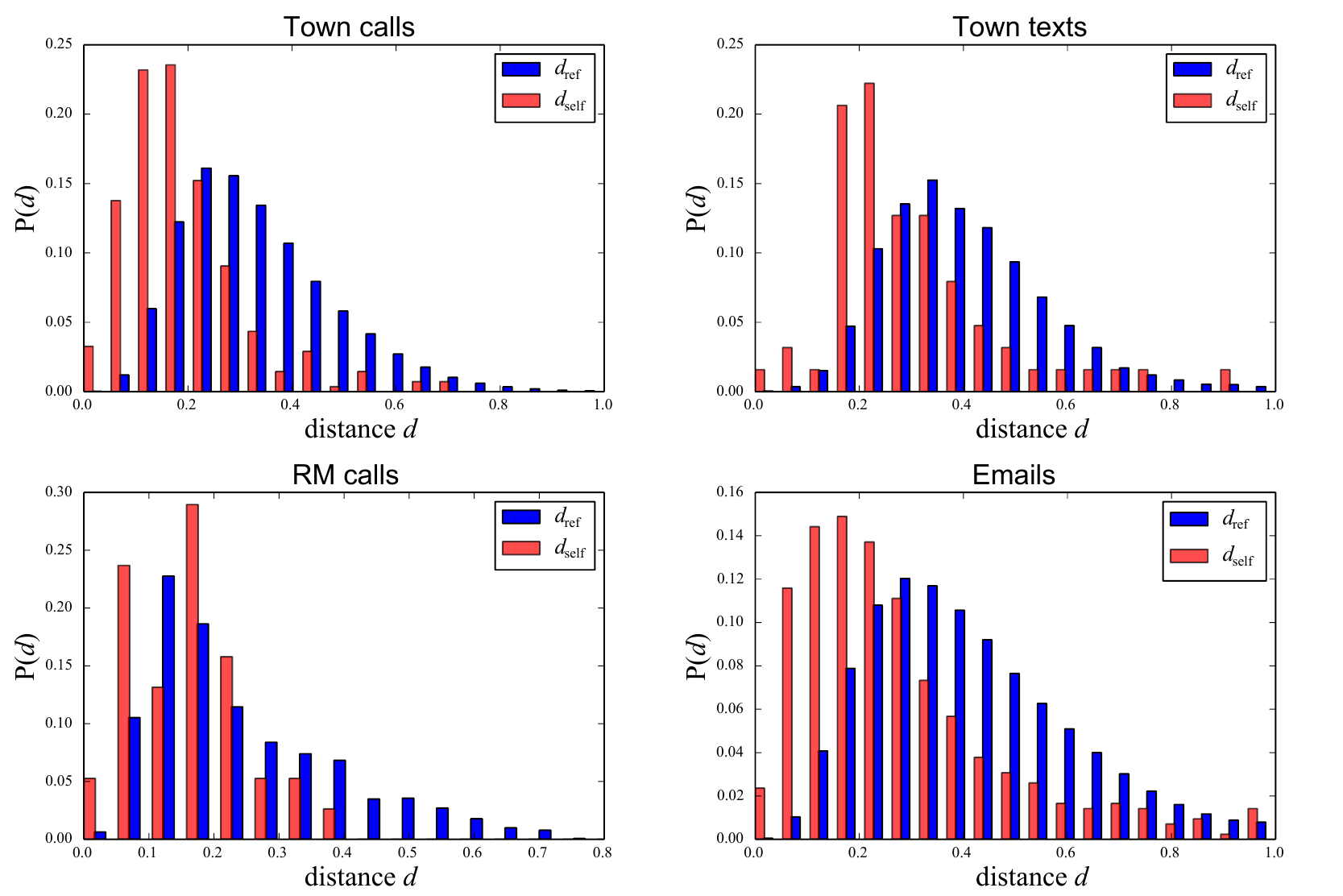}
\caption{Self and reference distances for daily patterns in our datasets. Self-distance measures the distance between one individual's daily patterns in two consecutive one-month intervals, whereas reference distances are computed between all pairs of individuals in a one-month interval.}
\label{fig:jsd}
\end{figure*}

Using the same methodology as~\cite{Aledavood2015} in order to study whether these daily patterns for each individual are persistent and thus characteristic for the individual, we divide the 8 weeks of data into two 4-week time intervals and use the Jensen-Shannon divergence to measure self and reference distances between patterns. A detailed explanation of these calculations can be found in the Methods section. The results are shown in Fig.~\ref{fig:jsd}. We observe an effect similar to the findings in ~\cite{Aledavood2015}: the daily patterns of individuals tend to be more similar to themselves in consecutive time intervals as compared to daily patterns of other individuals in the same time interval. This indicates that individuals have distinct daily patterns that retain their shapes in time.
In other words, Fig.~\ref{fig:jsd} shows that the individual differences seen in Fig.~\ref{fig:average_daily_patterns} are not just caused by random fluctuations: were fluctuations the reason for individual differences, each individual's patterns in consecutive intervals would be equally similar or dissimilar to those of everyone else. As self-distances are on average lower, this is clearly not the case.

\section{Discussion}

Circadian rhythms have deep roots in human physiology, driven by the environment in which we live. These patterns manifest themselves in different ways at the individual and aggregate levels. There are diurnal patterns that are only visible at the aggregate level in the overall frequencies of various phenomena that are rare or one-time events at the individual level:  
time of birth, heart attacks, suicides or committing unethical behaviour~\citep{refinetti2005, ruffieux1992,Kouchaki2014}. To the contrary, the daily rhythms that we have focussed on here originate at the level of individuals, where they manifest as time-dependent event rates of e.g.~digital communication.

What are the factors that determine an individual's daily rhythm as viewed through the lens of electronic records? %and what are the factors that give rise to the observed individual differences? 
The most obvious one is the sleep/wake cycle: we do not send emails or edit Wikipedia while asleep. This is known to be the central driver behind individual differences. First, individuals have different intrinsic chronotypes (morningness/eveningness tendencies~\citep{Kerkhof1985}). Second, the preferred duration of sleep also varies from one person to another~\citep{Blatter2007}. Third, besides these intrinsic factors, external forcing such as different work schedules also have an effect on the sleep/wake cycle~\citep{Taillard1999}. 

In addition to  differences in the sleep/wake cycle, our alertness and propensity to sleep are distinct for each individual and vary throughout the day. Naturally, individuals go on average through fairly similar cycles of wakefulness and sleepiness, which may explain the qualitatively similar features of aggregate-level daily patterns across different systems. At the level of individuals, however, there are important differences, which are reflected in the observed daily patterns in digital records. As an example, a tired person might be less likely to write an important email or edit a Wikipedia article. Likewise, in addition to these intrinsic alertness cycles, one's daily schedule (work, commuting, etc.) plays a role by imposing constraints on the times when it is possible to send emails or make calls. In terms of  daily patterns of telephone calls, things are more complicated, because every call involves two individuals---a caller and a recipient. When calling, one must consider social norms and the availability of the other party.

Understanding which of the factors discussed above dominate the digital daily cycles of individuals and give rise to individual differences and persistent circadian patterns is a task that requires further attention. While the persistence of daily patterns appears to indicate that the intrinsic components (chronotypes, alertness cycles) do play a major role~\citep{Aledavood2015}, external factors should also be of importance (see, e.g.,~\cite{Llorente2014}). Further, it will be necessary to study whether individuals bound by (strong) social ties tend to synchronise their communication and availability. 

While analysing digital records at the aggregate level can provide us invaluable population-level insights and help to replace or improve traditional survey or census methods~\citep{deville2014,toole2012}, studying the temporal fingerprints of individuals will unveil many new opportunities. As smartphones and other wearable devices are becoming ever more ubiquitous, they also increasingly provide high-velocity, high-volume data streams describing human behaviour~\citep{Torous2015_1}. This data-collection capability makes these devices excellent tools for research, particularly within health, psychology and medicine, since smartphones allow researchers to study individual behavioural patterns (``digital phenotypes'',~\citep{Jain2015,DigitalPhenotype2015}) and their changes over time~\citep{miller2012}. Monitoring an individual's digital behavioural patterns on different timescales is also an easy and inexpensive way for medical intervention, especially in the case of mental problems, where there are fewer biomarkers than for other types of disease. Data from smartphones have already been used to monitor the time evolution of different measures that are known to be indicative of behavioural changes in patients, which makes daily monitoring and early intervention possible ~\citep{Matthews2014,Torous2015_2,Saeb2015}. As an example, ~\cite{FaurholtJepsen2014} suggest that data from mobile phones can be used as objective measure of symptoms of bipolar disorder.

Because the sleep/wake cycle is a dominant feature of circadian patterns, Big Data describing the digital daily cycles of large numbers of individuals might prove to be highly useful for sleep research. However, obtaining an accurate picture of the sleep times of individuals requires solving several non-trivial problems. While one does not send emails when asleep, emails are not necessarily a reliable proxy for awake-time; it is possible to be awake and not send emails. In this sense inferring the actual times of sleep from electronic records is challenging. This problem is made more severe by the ubiquitous burstiness in human dynamics~\citep{Barabasi2005,Karsai2011,Miritello2011}: broadly distributed inter-event times make the times from last observation to bed time (or from wake-up to first observation) highly unpredictable. Nevertheless, we believe that this is an important direction for future research. 

Finally, a particularly promising source of data comes from large dedicated cell-phone based data collection efforts, focusing on collecting multiplex (face-to-face, telecommunication, online social networks) network data in a large, densely connected populations, e.g.,~\cite{stopczynski2014measuring}. Data from a single communication channel can be too sparse and noisy for obtaining accurate daily patterns; here, having a multiplex dataset can provide a great advantage since one can combine information from all data-channels to form a much more comprehensive picture of the activity of each person (e.g.~for studying sleeping patterns). Furthermore, if the participants of the dataset are densely connected through social ties, it is also possible to investigate the significance of and correlations between the activity patterns of close personal relations using such a dataset. Finally, a dataset of this nature may function as a kind of ``rosetta stone'', helping researchers determine the biases of each electronic dataset, and allowing us to understand to which extent telecommunication data or Twitter datasets with hundreds of millions of active users can be used to study the daily cycles of individuals.

\section{Methods}
\subsection{Data filtering}
We have used 8-week time slices of all datasets. Filters have been applied to remove users who are inactive or whose activity is too low for producing meaningful information on daily patterns. In Table~\ref{table_data}, the total number of participants means the total number of users who have at least one event during the study period of 8 weeks. For plotting aggregate-level patterns (Fig.~\ref{fig:aggregated} and Fig.~\ref{fig:wednesday}), we have used data from all participants. The column ``Active users'' in the table represents the number of users who have at least one event per day on average (minimum 56 events in total); these have been used for calculating average daily patterns (Fig.~\ref{fig:average_daily_patterns}). For measuring persistence of daily patterns and calculating Jensen-Shannon divergence, we used a subset of active users who have at least one event in each of the two time intervals of 4 weeks. 

\subsection{Self and reference distances}
In order to quantify the level of persistence of daily patterns for individuals, we compare the daily patterns of each ego for two consecutive 4-week time intervals. For this, we use the Jensen-Shannon divergence (JSD) and measure the distance of the daily patterns viewed as two probability distributions ($P_1$ and $P_2$). The JSD is calculated as follows:
$JSD(P_1,P_2) = H(\frac{1}{2}P_1 + \frac{1}{2}P_2)- \frac{1}{2}[H(P_1)-H(P_2) ]$, where $P_i = p(h)$ and $p(h)$ is the fraction of calls at each hour, $i=1,2$ indicates the time interval, and $H(P)= -\sum p(h) \log p(h)$ is the Shannon entropy. In order to compare these self-distances against a reference, we calculate a set of reference distances $d_\mathrm{ref}$ as the distances between the daily patterns of each ego and all other egos in the same time interval.

\section*{Disclosure/Conflict-of-Interest Statement}

%Frontiers follows the recommendations by the International Committee of Medical Journal Editors (http://www.icmje.org/ethical_4conflicts.html) which require that all financial, commercial or other relationships that might be perceived by the academic community as representing a potential conflict of interest must be disclosed. If no such relationship exists, authors will be asked to declare that the research was conducted in the absence of any commercial or financial relationships that could be construed as a potential conflict of interest. When disclosing the potential conflict of interest, the authors need to address the following points:
%‚Ä¢	Did you or your institution at any time receive payment or services from a third party for any aspect of the submitted work?
%‚Ä¢	Please declare financial relationships with entities that could be perceived to influence, or that give the appearance of potentially influencing, what you wrote in the submitted work.
%‚Ä¢	Please declare patents and copyrights, whether pending, issued, licensed and/or receiving royalties relevant to the work.
%‚Ä¢	Please state other relationships or activities that readers could perceive to have influenced, or that give the appearance of potentially influencing, what you wrote in the submitted work.

The authors declare that the research was conducted in the absence of any commercial or financial relationships that could be construed as a potential conflict of interest.

\section*{Author Contributions}
%When determining authorship the following criteria should be observed:
%‚Ä¢	Substantial contributions to the conception or design of the work; or the acquisition, analysis, or interpretation of data for the work; AND
%‚Ä¢	Drafting the work or revising it critically for important intellectual content; AND
%‚Ä¢	Final approval of the version to be published ; AND
%‚Ä¢	Agreement to be accountable for all aspects of the work in ensuring that questions related to the accuracy or integrity of any part of the work are appropriately investigated and resolved.
%Contributors who meet fewer than all 4 of the above criteria for authorship should not be listed as authors, but they should be acknowledged. (http://www.icmje.org/roles_a.html)

TA, SL, and JS designed research. TA analysed the data. TA, SL, and JS wrote the paper.

\section*{Acknowledgments}

\textit{Funding\textcolon} TA and JS acknowledge financial support from the Academy of Finland, project No.~260427. TA thanks Richard Darst for useful discussions.

% \section*{TO BE REMOVED}
% The article for Frontiers is included in a "research topic"
% so according to this page http://www.frontiersin.org/about/RTGuidelines
% we are free to pick the format. And therefore, we pick "original research" which has to be 
% maximum 12,000 words and 15 figures.
% Also:

% "For Original Research Articles, it is recommended to organize your manuscript in the following sections or their equivalents for your field:
% Introduction
% Succinct, with no subheadings.
% Material and Methods
% This section may be divided by subheadings. This section should contain sufficient detail so that when read in conjunction with cited references, all procedures can be repeated.
% Results
% This section may be divided by subheadings. Footnotes should not be used and have to be transferred into the main text.
% Discussion
% This section may be divided by subheadings. Discussions should cover the key findings of the study: discuss any prior art related to the subject so to place the novelty of the discovery in the appropriate context; discuss the potential short-comings and limitations on their interpretations; discuss their integration into the current understanding of the problem and how this advances the current views; speculate on the future direction of the research and freely postulate theories that could be tested in the future."

%\bibliographystyle{frontiersinSCNS_ENG_HUMS} % for Science, Engineering and Humanities and Social Sciences articles, for Humanities and Social Sciences articles please include page numbers in the in-text citations
\bibliographystyle{frontiersinHLTH_FPHY} % for Health and Physics articles
\bibliography{frontiers_refs}

\end{document}